\documentclass[10pt,twocolumn,superscriptaddress,english,10pt,showpacs,floatfix]{revtex4-1}
\usepackage{amsthm}
\usepackage{amsmath}
\usepackage{amssymb}
\usepackage{mathdots}
\usepackage{graphicx}
\usepackage[normalem]{ulem}
\usepackage{array}
\makeatletter
\@ifundefined{textcolor}{}
{%
 \definecolor{BLACK}{gray}{0}
 \definecolor{WHITE}{gray}{1}
 \definecolor{RED}{rgb}{1,0,0}
 \definecolor{GREEN}{rgb}{0,1,0}
 \definecolor{BLUE}{rgb}{0,0,1}
 \definecolor{CYAN}{cmyk}{1,0,0,0}
 \definecolor{MAGENTA}{cmyk}{0,1,0,0}
 \definecolor{YELLOW}{cmyk}{0,0,1,0}
}

\usepackage{braket}
\usepackage[backref=true,
bookmarksnumbered=true,
bookmarks=true,
bookmarksopen=true,
colorlinks=true,
citecolor=blue,
linkcolor=blue,
anchorcolor=green,
urlcolor=blue,unicode=false]{hyperref}
\renewcommand{\v}[1]{\ensuremath{\mathbf{#1}}} % for vectors
 
\newcommand{\abs}[1]{\left| #1 \right|} % for absolute value
\let\baraccent=\= % rename builtin command \= to \baraccent
\renewcommand{\=}[1]{\stackrel{#1}{=}} % for putting numbers above =

\newcolumntype{L}[1]{>{\raggedright\let\newline\\\arraybackslash\hspace{0pt}}m{#1}}
\newcolumntype{C}[1]{>{\centering\let\newline\\\arraybackslash\hspace{0pt}}m{#1}}
\newcolumntype{R}[1]{>{\raggedleft\let\newline\\\arraybackslash\hspace{0pt}}m{#1}}

\makeatother

\usepackage{babel}

\begin{document}
\title{Proximity-induced Majorana hinge modes in antiferromagnetic topological insulators}

\author{Yang Peng}
%\email{yangpeng@caltech.edu}
\affiliation{Institute of Quantum Information and Matter and Department of Physics,California Institute of Technology, Pasadena, CA 91125, USA}
\affiliation{Walter Burke Institute for Theoretical Physics, California Institute of Technology, Pasadena, CA 91125, USA}
\author{Yong Xu}
%\email{yongxu@mail.tsinghua.edu.cn}
\affiliation{State Key Laboratory of Low Dimensional Quantum Physics, Department of Physics, Tsinghua University,
Beijing 100084, People’s Republic of China}
\affiliation{Collaborative Innovation Center of Quantum Matter, Beijing 100084, People’s Republic of China}
\affiliation{RIKEN Center for Emergent Matter Science (CEMS), Wako, Saitama 351-0198, Japan}

\begin{abstract}
	We propose a realization of chiral Majorana modes propagating on the hinges of a 3D antiferromagnetic topological insulator, which
	was recently theoretically predicted and experimentally confirmed in the tetradymite-type $\mathrm{MnBi_2Te_4}$-related ternary chalgogenides. 
	These materials consist of ferromagnetically ordered 2D layers, whose magnetization direction alternates between neighboring layers, forming
	an antiferromagnetic order. Besides surfaces with a magnetic gap, there also exsist gapless surfaces with a single Dirac cone,
	which can be gapped out when proximity coupled to an $s$-wave superconductor. On the sharing edges between the two types of
	gapped surfaces, the chiral Majorana modes emerge.  We further propose experimental signatures of these Majoana hinge modes
	in terms of two-terminal conductance measurements. 
\end{abstract}

\maketitle
\section{Introduction}
Majorana edge mode, appearing as a gapless excitation on the boundary of a topological superconductor (TSC), 
has attracted a lot of attention because of its unusual property in analogy to 
the theoretically proposed Majorana fermion in particle physics, which 
is its own antiparticle \cite{Hasan2010,Qi2011,Alicea2012,Beenakker2013,Aguado2017}.
The zero dimensional version of Majorana modes are zero-energy
excitations localized at the ends of a 1D TSC, and thus give rise to degenerate many-body ground states,   
which can be used as nonlocal qubits and memory for quantum computing \cite{Kitaev2003,Nayak2008,Aasen2016}. 
Engineering Majorana zero modes in a variety of systems has been proposed theoretically
\cite{Fu2008,Lutchyn2010,Oreg2010,Nadj-Perge2013,Pientka2013,Peng2015} and
tested experimentally \cite{Mourik2012,Das2012,Churchill2013,Deng2012,Finck2013,Nadj-Perge2014,Ruby2015,Pawlak2016,Deng2016,Albrecht2016,Ruby2017,Gul2018}.

The 1D chiral Majorana mode (CMM) is a unidirectionally propagating mode appearing on the boundary of a 2D 
$p\pm ip$ chiral superconductor \cite{Reed2000}, which has a full pairing gap in the bulk and can be regarded as 
the superconducting analog of a Chern insulator. The propagation of the 1D CMMs  
has been shown in Ref.~\cite{Lian2017} to give rise to the similar qubit operations as Majorana zero modes do,
enabling performing quantum computation with CMMs.

On the experimental side, the CMMs were proposed to be realized in a
heterostructure comprising a quantum anomalous Hall insulator (QAHI)
and an $s$-wave superconductor \cite{Qi2010,Chung2011,Strubi2011,Wang2015}. 
Based on this proposal, it was reported in a recent experiment \cite{He2017}
that the CMM was observed via a transport measurement of $e^2/2h$ conductance plateau 
in a QAHI-TSC-QAHI junction formed with a Cr-doped $\rm (Bi,Sb)_2Te_3$ thin films in proximity with
a Nb superconductor. 

However, the interpretation of this conductance plateau as a signature
for the presence of CMMs is under debate.
In this experiment, an external magnetic field is required to tune the thin film into 
a magnetization reversal stage, when the system is near a QAHI-normal insulator 
phase transition \cite{He2017}. It is expected that the system in this
magnetization reversal stage is extremely inhomogeneous, which
leads to alternative explanations of the conductance plateau 
under strong disorders without CMMs \cite{Ji2018,Huang2018,Lian2018}.
Hence, it is desirable to have a platform hosting CMMs without suffering from
sample inhomogeneity, as it happened in the Cr-doped $\rm (Bi,Sb)_2Te_3$ thin films.

Very recently, a new 3D bulk material, the ``antiferromagnetic topological insulator'' (AFMTI) \cite{Mong2010},
was predicted theoretically \cite{Li2018,Zhang2018,Otrokov2018} and
confirmed experimentally \cite{Otrokov2018,Gong2018,Chowdhury2019,Lee2018,Yan2019} in the tetradymite-type $\mathrm{MnBi_2Te_4}$-related
ternary chalgogenides ($\mathrm{MB_2T_4}$: M = transition-metal or rare earth element, B = Bi or Sb, T = Te, Se or S).
This material has both topological nontrivial band structure, as well as intrinsic magnetic order, 
namely an inter-layer antiferromagnet with perpendicular magnetic anisotropy. 
Because of the intrinsic magnetization, the magnetic gap created 
is expected to be large and uniform, and would presumably raise the 
observable temperature of the quantum anomalous Hall effect \cite{Tokura2019}. 
It is worth mentioning that the magnetic gap was already observed
by ARPES \cite{Otrokov2018}, while a finite anomalous Hall effect
was also measured by transport experiments \cite{Gong2018,Lee2018}. 

\begin{figure}[t]
	\centering
	\includegraphics[width=0.45\textwidth]{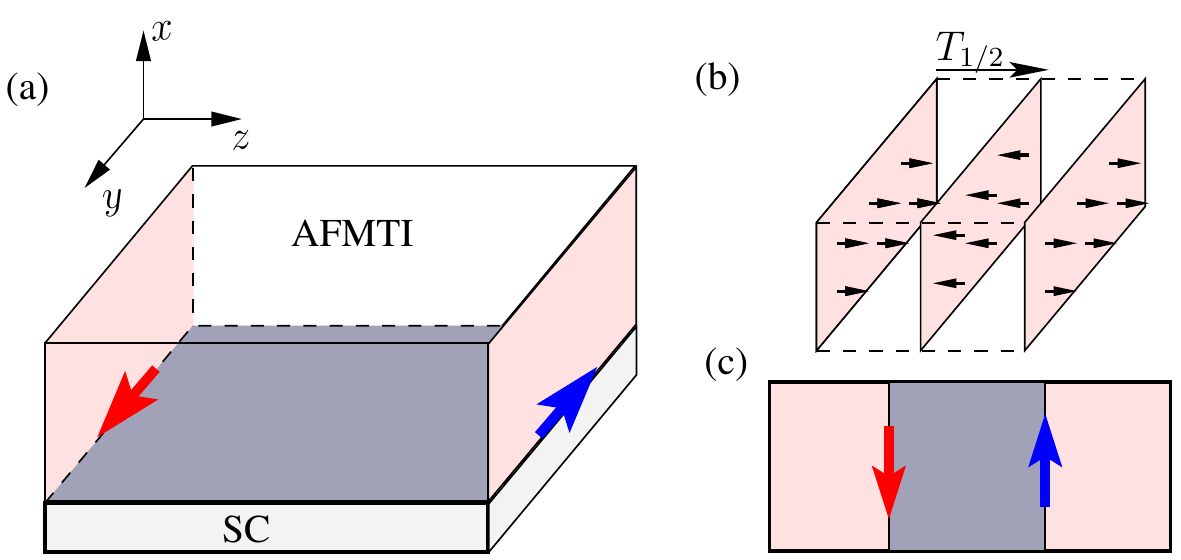}
	\caption{\label{fig:setup} (a) Majorana hinge modes (blue and red arrows) at the edge of the interface (in grey) between an AFMTI and an $s$-wave SC. The
		antiferromagnetic ordering and the magnetization direction are both assumed to be along $z$ direction. We also assume the left and right surfaces (in pink)
		to have opposite
magnetization. (b) The AFMTI can be regarded as magnetic layers, which are ferromagnetically ordered within each layer, and antiferromagnetically ordered between layers. 
(c) Effective description of the left-bottom-right surfaces of the AFMTI, in which the hinge modes appears at the domain wall between magnetic gapped (pink) 
and superconducting gapped (grey) regions. 
}
\end{figure}

Given the very nice properties of the AFMTI, and the experimental feasibility of growing clean bulk samples, 
one may ask if one can use the AFMTI as a platform to create robust CMMs?
In this manuscript, we provide a definite answer to this question.
Particularly, if an $s$-wave superconductor is coupled 
to the AFMTI surfaces with zero net magnetization, which is parallel to the antiferromagnetic direction, 
a superconducting gap would be induced on these surfaces. 
We propose that the CMMs can be created on the hinges of the AFMTI (Fig.~\ref{fig:setup}(a)),
which is shared by magnetically gapped surfaces and the superconducting gapped surfaces. 

The advantage of our proposal is that the CMMs are expected to be observed
at a higher temperature, thanks to the intrinsic magnetic order in the bulk crystal of the AFMTI.
Since no additional magnetic proximity/fields is required,
the complication when the magnetism and superconductivity are spatially overlapping,
as in those 2D platforms (Cr-doped $\rm (Bi,Sb)_2Te_3$ thin films), can be avoided.
More intriguingly, it was shown that by manipulating the number parity of layers of the AFMTI
such as $\mathrm{MnBi_2Te_4}$, one is able to switch the system between a quantum anomalous Hall state
and an axion insulator state \cite{Li2018}. We will demonstrate in our manuscript that
the propagating direction of the CMMs can also be controlled, using the idea of 
changing the number of AFMTI layers. This would make  
our proposal more flexible in designing complicate networks of CMMs 
than the one in Ref.~\cite{Lado2018}, in which a trivial antiferromagnetic insulator was used. 

It is worth mentioning that searching for systems with hinge modes is one of the active directions
in the field of higher-order topological insulators and
superconductors \cite{Benalcazar2017s,Peng2017,Langbehn2017,Benalcazar2017,Song2017,Schindler2018, Wang2018, Geier2018}.
The AFMTI with proximity superconducting gaps proposed in this work can be regarded as an ``extrinsic'' second-order
topological superconductor.
By ``extrinsic'', it means the hinge modes depend not only on the bulk band structure, 
but also on the boundaries, i.e. on the lattice termination.
Another example of ``extrinsic'' second-order topological systems is a 
three-dimensional time-reversal invariant topological insulator (TRITI)
placed in a magnetic field in a generic direction, such that there is a finite magnetic flux through all surfaces
\cite{Sitte2012,Zhang2013}. 
These systems should be distinc from the ``intrinsic'' second-order topological systems
with corner or hinge states, which do not depend on lattice termination. These systems 
require additional approximate crystalline symmetries, including reflection symmetry
\cite{Langbehn2017,Schindler2018}, as well as rotation symmetries \cite{Song2017,Schindler2018}.
(Please refer to Ref.~\cite{Geier2018} for a more thorough discussion on the differences between 
``extrinsic'' and ``intrinsic'' higher-order topological systems.)

The rest of the paper is organized as follows.
We will first recall the basic properties of an AFMTI in Sec.~\ref{sec:AFMTI}, 
and demonstrate that proximity superconductivity does open a gap
on the AFMTI surfaces with zero net magnetization, despite the broken time-reversal
symmetry. In Sec.~\ref{sec:MHM}, we show that the CMMs appear
at the hinges of the AFMTI, which are shared by the
magnetically gapped and superconducting gapped surfaces. 
In Sec.~\ref{sec:experiments}, 
we discuss experimental signatures of these CMMs in terms of transport measurements. 
Finally, we conclude in Sec.~\ref{sec:conclusion}.

\section{Antiferromagnetic topological insulator \label{sec:AFMTI}} 
The AFMTI can be viewed as a TRITI 
with additional staggered time-reversal breaking terms \cite{Mong2010},
such as antiferromagnetically ordered layers of magnetic moments as shown in Fig.~\ref{fig:setup}(b).
This picture was recently demonstrated by \textit{ab initio} calculations of materials such as $\mathrm{MnBi_2Te_4}$ \cite{Li2018,Zhang2018}. 
In $\mathrm{MnBi_2Te_4}$, the staggered magnetic potential that breaks the time-reversal symmetry is generated by the $\mathrm{Mn}$ atoms,
while topological states are introduced by the Bi-Te layers same as in $\mathrm{Bi_2Te_3}$ \cite{Zhang2009}.
It was reported that the states close to the Fermi level are $p$-bands of Bi/Te, and the $\mathrm{Mn}$ $d$-bands are far away from the 
band gap with an extremely large exchange splitting ($>7\mathrm{eV}$) \cite{Li2018}.

Despite the broken time-reversal $\Theta$ symmetry, the AFMTI
can have topological nontrivial features,
because the symmetry $\mathcal{S}=\Theta T_{1/2}$ is preserved \cite{Mong2010}, where $T_{1/2}$ is a primitive
lattice translation symmetry that itself is broken by the antiferromagnetic order. 
One important difference between the AFMTI and the TRITI is that not all surfaces are gapless.
Indeed, the surfaces are gapless only when they preserve the bulk symmetry $\mathcal{S}$,
and these surfaces are of type A (antiferromagnetic).
There are other surfaces which break the $\mathcal{S}$ symmetry,
and are of type F (ferromagnetic).
As shown in Fig.~\ref{fig:setup}(a), 
the top, bottom, front and back surfaces of the AFMTI are of type A
whereas the left and right are type F surfaces. 

It has been shown in Ref.~\cite{Fu2008} that  
the gapless surface states of a time-reversal invariant topological insulator (TRITI),
can be gapped out by either breaking the time-reversal symmetry, 
when the surface is coupled to an magnetic insulator,
or by superconductivity, when the surface is coupled to an $s$-wave superconductor.
At the domain wall between the two gapped regions, the CMMs will emerge. 

In an AFMTI, the gapped surface state on a type F surface
is very similar to the one on a TRITI surface coupled to a
magnetic insulator. On the other hand, the type A surface
has zero net magnetization and hosts a single gapless
Dirac cone. Then the natural
question to ask is that is it possible to gap it out by coupling
to an $s$-wave superconductor? If so,  on the sharing hinges between the two types of gapped surfaces, the CMMs
should appear as indicated in red and blue arrows in Figs.~\ref{fig:setup}(a,c).

Although both a TRITI surface and a type A surface of an AFMTI support a single gapless Dirac cone, 
the gapless nature of these two kinds of surfaces are not expected to be the same.
While the former is protected by the physical time-reversal symmetry, the latter
is protected by the composite symmetry $\mathcal{S}$. 
It is unclear whether introducing $s$-wave superconductivity can 
also gap out the type A surface of an AFMTI, in the same way as
it does on a TRITI surface, despite the broken physical time-reversal symmetry.

In the following, we will first demonstrate analytically that a generic type A surface
of an AFMTI can indeed be gapped out by $s$-wave superconductivity. 
We will then numerically verify this on a concrete tight-binding model on
a cubic lattice \cite{Mong2010}, which captures all essential (topological) properties of a realistic AFMTI.

\subsection{Generic considerations}
\subsubsection{Nature of gapless type A surfaces}
For a generic surface states of a TRITI, it can be gapped out immediately as the time-reversal 
symmetry is broken, while this is not the case for a type A surface of an AFMTI. 
To further understand the difference between the latter and a generic TRITI surface with time-reversal breaking
potentials, let us consider an AFMTI with a Bloch Hamiltonian written as $\mathcal{H}(\v{k})$, 
where $\v{k}$ is a point in the 3D Brillouin zone. 
Recall that the AFMTI can be viewed as a TRITI with a staggered time-reversal breaking field 
switched on, without closing the bulk gap. Thus, the AFMTI acquires a sublattice structure with 
opposite time-reveral breaking fields on the two types of sublattices within a unit cell.  
If we introduce a set of Pauli matrices $\mu_{j}$, $j=x,y,z$, for this sublattice degree of freedom, 
we can write the Bloch Hamiltonian of the AFMTI as
\begin{equation}
\mathcal{H}(\v{k}) = \mathcal{H}_0(\v{k}) + V\mu_z,
\end{equation}
where $\mathcal{H}_0(\v{k})$ is the Bloch Hamiltonian for a TRITI, and the potential $V$ breaks the physical time-reversal symmetry,
namely $\{\Theta,V\}=0$.

Let us denote the lattice vectors as $\v{a}_j$ for $j=1,2,3$, 
and introduce $k_j=\v{k}\cdot\v{a}_j$ as the coordinate in the Brillouin zone. 
Without loss of generality, we can assume the AFMTI acquires an antiferromagnetic order along $\v{a}_3$. 
Since the AFMTI has a symmetry $\mathcal{S}=\Theta T_{1/2}$, where $T_{1/2}$ is a half lattice translation
along $\v{a}_3$, we have that the effective 2D Hamiltonian 
\begin{equation}
\mathcal{H}_{\mathrm{eff}}(\v{k}_{12}) = \left.\mathcal{H}(\v{k})\right|_{k_3=0}
\end{equation}
has an effecitve time-reversal symmetry realized by $S = \mu_x\Theta$ as
\begin{equation}
S\mathcal{H}_{\rm eff}(\v{k}_{12}) = \mathcal{H}_{\rm eff}(-\v{k}_{12})S,
\end{equation}
with $\v{k}_{12}=(k_1,k_2)$.
Because of this effective time-reversal symmetry, this 2D system acquires a $\mathbb{Z}_2$
topological classification, as in a quantum-spin-Hall insulator \cite{Kane2005}.

To understand the robustness of the gapless Dirac cone on the type A surfaces,
let us first consider $V=0$, when $\mathcal{H}(\v{k})=\mathcal{H}_0(\v{k})$ is
a TRITI with gapless surface states.
This means $\mathcal{H}_{\mathrm{eff}}(\v{k}_{12})$ is in a nontrivial phase
that supports helical boundary modes.

Indeed, we have $[\mathcal{H}_{\mathrm{eff}}(\v{k}_{12}),\mu_x]=0$ when $V=0$,
which means $\mathcal{H}_{\mathrm{eff}}(\v{k}_{12})$ can be block diagonalized
into two blocks according to the eigenvalues $\pm 1$ of $\mu_x$. Due to $\mathbb{Z}_2$
topological classification of $\mathcal{H}_{\rm eff}$, only one of the two blocks
supports a pair of helical modes, while the other block is trivially gapped. 

Let us consider the system is only periodic along $\v{a}_{2}$ and open along $\v{a}_1$,
we can denote the effecive Hamiltoinian of the gapless block describing the edge along 
$\v{a}_2$ by
\begin{equation}
\mathcal{H}_{\rm eff}^{\rm edge}(k_2;V=0) = k_2\Gamma,
\end{equation}
where $\Gamma$ is a two-by-two Hermitian matrix with eigenvalues $1$ and $-1$, 
after rescaling the Fermi velocity of the helical modes to unity. 
We further denote the gapped Hamiltonian of the other block as $h_{\rm gap}(k_2)$.

When we switch on the time-reversal breaking potential $V$,
a coupling is introduced between the two blocks corresponding to the $\pm 1$ eigenvalues of $\mu_x$.
The low-energy Hamiltonian of the gapless sector can be written as 
\begin{equation}
\mathcal{H}_{\rm eff}^{\rm edge}(k_2) = k_2\Gamma + V h_{\rm gap}(k_2)^{-1} V^\dagger.
\end{equation}
Since $\{\Theta,V\}=0$ and $h_{\rm gap}(k_2)$ respects the time-reversal symmetry $\Theta$,
$Vh_{\rm gap}^{-1}V^\dagger$ will also respect $\Theta$ and the helical edge mode cannot be gapped out.
This leads to the robustness of the gapless states on the type A surfaces.

\subsubsection{Proximity induced superconducting gap}
What happens if we introduce $s$-wave superconductivity on the type A surfaces?
Let us keep the open boundary condition along $\v{a}_1$ as we did above, 
and couple the type A surface of the AFMTI parallel to $\v{a}_2$ and $\v{a}_3$,
to an $s$-wave superconductor, such as Nb.
We still keep the periodic boundary condition along $\v{a}_2$ and $\v{a}_3$.

Due to superconducting proximity effect, an intraorbital $s$-wave pairing 
$\Delta(j) = \braket{c_{A,\v{k}_{23}\uparrow\beta}^\dagger(j) c_{A,-\v{k}_{23}\downarrow\beta}^\dagger(j)} 
= \braket{c_{B,\v{k}_{23}\uparrow\beta}^\dagger(j) c_{B,\v{-k}_{23}\downarrow\beta}^\dagger(j)}$
can be created.  
Here $c_{X,\v{k}_{23}(j)\sigma\beta}^\dagger$ ($X$=$A$,$B$, $\sigma=\uparrow,\downarrow$)
creates an electron at either of the two sublattices $A$ or $B$, 
in orbital $\beta$, with momentum $\v{k}_{23} = (k_2,k_3)$ and spin $\sigma$,
at the coordinate $j$ along $\v{a}_1$. 
The induced pairing strength $\Delta(j)$ decays exponentially into the bulk along $\v{a}_1$,
namely $\Delta(j) = \Delta_0 \exp(-j/\xi)$, with a localization length $\xi$.

The Bloch Bogoliubov--de Gennes (BdG) Hamiltonian for the AFMTI
with proximity induced $s$-wave pairing has the following form
\begin{equation}
	\mathcal{H}_\mathrm{BdG}(\v{k}_{23}) = \left(\begin{array}{cc}
\mathcal{H}(\boldsymbol{k}_{23}) & -i\Delta\sigma_{y}\\
i\Delta\sigma_{y} & -\mathcal{H}(-\boldsymbol{k}_{23})^{*}
\end{array}\right),
\label{eq:BdG}
\end{equation}
where we have introduced the Pauli matrices $\sigma_{x,y,z}$ for the spin degree of freedom, 
and have chosen the time-reversal operation as $\Theta = -i\sigma_y\mathcal{K}$, with complex conjugation $\mathcal{K}$.
The matrix $\mathcal{H}(\v{k}_{23})$ above corresponds to the Hamiltonian of the AFMTI,
with real-space representation used along $\v{a}_1$, and is local in $k_2,k_3$ 
due to the periodic boundary condition.
The pairing matrix $\Delta$ is diagonal in real space coordinate along $\v{a}_1$, and is
local in $k_2, k_3$ as well. 

Similar to the analysis of the gapless modes without superconductivity, 
one can focus at $k_3=0$ and demonstrate that the gapless modes can indeed be gapped out
by $s$-wave superconductivity.
We will denote
\begin{equation}
\mathcal{H}_{\rm BdGeff}(k_2) = \left.\mathcal{H}_\mathrm{BdG}(\v{k}_{23})\right|_{k_3 = 0},
\end{equation}
which corresponds to 
substituting $\mathcal{H}(\v{k}_{23})$ by $\mathcal{H}_{\rm eff}$ in Eq.~(\ref{eq:BdG}).

Using the same apporach in analyzing the
helical edge states of $\mathcal{H}_{\rm eff}$, we first set the time-reversal
breaking field $V$ to zero, which gives rise to $[\mathcal{H}_{\rm BdGeff}(k_2),\mu_x]=0$. 
Hence, $\mathcal{H}_{\rm BdGeff}$, same as $\mathcal{H}_{\rm eff}$, can be block diagonalized
into two blocks corresponding to the eigenvalues $\pm 1$ of $\mu_{x}$. 
We can further make use of the effective time-reversal symmetry $S = -i\mu_x\sigma_y$ for $\mathcal{H}_{\rm eff}$, 
and write the low-energy edge BdG Hamiltonian as
\begin{equation}
\mathcal{H}_{\rm BdGeff}^{\rm edge}(k_2;V=0) = (k_2\Gamma\tau_z + \Delta\tau_y) \oplus (h_{\rm gap}(k_2)\tau_y + \Delta
\tau_y),
\label{eq:BdGeffedge}
\end{equation}
where $\tau_{x,y,z}$ are Pauli matrices in the Nambu space. 

We see the edge along $\v{a}_2$ are indeed gapped out when superconductivity is introduced. Particularly, the
nature of the gaps in the two sectors, which correspond to the $\pm 1$ eigenvalues of $\mu_x$, 
are very different. The first sector (first term in Eq.~(\ref{eq:BdGeffedge})) contains a pair of
superconducting gapped states, originating from the original gapless helical modes.
While in the other sector, the gap is much larger and its nature is the same as the one in a trivial insulator, 
assuming the gap of $h_{\rm gap}(k_2)$ is much larger than the superconducting pairing strength $\Delta$.

When we switch on the time-reversal breaking field $V$ before closing the bulk gap, 
a coupling is introduced between the two sectors. However, the low-energy theory of the gapped
edge does not change and is still given by the superconducting gapped helical modes, namely
\begin{equation} 
\mathcal{H}_{\rm BdGeff}^{\rm edge}(k_2) = k_2\Gamma\tau_z + V(h_{\rm gap}(k_2)\tau_y + \Delta
\tau_y)^{-1} V^{\dagger}+\Delta\tau_y,
\end{equation}
where the additional term $V(h_{\rm gap}(k_2)\tau_y + \Delta\tau_y)^{-1} V^{\dagger}$ 
respects the time-reversal symmetry $\Theta$.

Thus, we see that the type A surface of an AFMTI can indeed be gapped out by $s$-wave superconductivity, 
due to proximity effect. Moreover, the low-energy nature of this gapped surface states is exactly the 
same as the one obtained from gapping out a TRITI surface by superconductivity. 

It is worth mentioning that in the above discussion, 
we have assumed the superconducting proximity effect is induced by local electron
tunneling between the AFMTI type-A surface and the s-wave superconductor. Thus, the proximity-induced
superconductivity is then well approximated by the on-site s-wave pairing amplitude which 
decays exponentially into the bulk \cite{Peng2015}. 
In fact, the accurate profile of the induced superconducting pairing potential near the type-A surface
will depend on the strength of the proximity coupling (interface quality, lattice mismatch
etc.), as well as the superconducting pairing potential of the bulk superconductor.
These require much more sophisticated ab initio modeling involving realistic lattice structures and material parameters, 
which will be investigated in the future. 

\subsubsection{Majorana Modes}
In the previous discussion, we have shown that the type-A surface with superconductivity, and the type-F
surface in an AFMTI resemble the two gapped regions with different topology in
the 2D Fu-Kane model.  On the other hand, we know that the 3D second-order topological
insulators follow the same topological classification of the 2D (first-order) topological insulator, 
in the sense that a 3D second-order topological insulator can be regarded as 
gluing a 2D topological insulator on its 2D boundaries \cite{Peng2017,Langbehn2017}.

Hence, we expect the CMMs appear on the sharing hinges between the two types of gapped surfaces
of the AFMTI, due to the 
above mentioned correspondence to a 2D Fu-Kane model (illustrated in Fig.~\ref{fig:setup}(c)).

In the following, we will consider a concrete tight-binding model \cite{Mong2010}, 
and explicitly verify the results obtained from the above general discussion. 

\subsection{Tight-binding model}
The tight-binding model is constructed from a four-band TRITI model defined on a cubic lattice (lattice constant equals to 1)
with the following Bloch Hamiltonian \cite{Hosur2010}
\begin{equation}
	\mathcal{H}_{\mathrm{TI}}(k_x,k_y,k_z) = m\rho_z + \sum_{j=x,y,z}( t\cos k_j\rho_z + \lambda\sin k_{j}\sigma_j\rho_x),
\end{equation}
where $\sigma_{j}$ and $\rho_{j}$ ($j=x,y,z$) are two sets of Pauli matrices for spin and orbital degree of freedom.
The time-reversal symmetry in this system is realized by $\Theta = -i\sigma_y\mathcal{K}$, with complex conjugation $\mathcal{K}$.
Note that the system is a strong topological insulator for $\abs{m}\in(\abs{t},3\abs{t})$ with finite spin-orbit coupling
($\lambda\neq 0$).
To have an AFMTI, we further introduce a staggered time-reversal-breaking field
alternating between $V$ and $-V$ in neighboring layers
along the $(\bar{1}\bar{1}1)$ direction, where $\{\Theta,V\}=0$.

In the antiferromagnetic state, the unit cell contains two sublattices $A$ and $B$,
with staggered potential $V$ and $-V$, respectively.
Let us choose $A$ and $B$ sit at positions $(0,0,0)$ and $(0,0,1)$, with respect to 
the original cubic lattice vectors $\hat{\v{x}},\hat{\v{y}}, \hat{\v{z}}$.
We can then define the new basis vectors  $\v{a}_1 = \hat{\v{x}} + \hat{\v{z}}$,
$\v{a}_2 = \hat{\v{y}} + \hat{\v{z}}$, and $\v{a}_3 = 2\hat{\v{z}}$ 
for the enlarged unit cell after introducing the staggered exchange field.  

The onsite potentials on the $A$ and $B$ sites are
\begin{equation}
	H_A = m\rho_z + V,\quad H_B = m\rho_z - V,
\end{equation}
respectively. The hopping terms 
\begin{gather}
	T_{\pm x} = (t\rho_z \pm i\lambda\rho_x\sigma_x)/2 \\
	T_{\pm y} = (t\rho_z \pm i\lambda\rho_x\sigma_y)/2 \\
	T_{\pm z} = (t\rho_z \pm i\lambda\rho_x\sigma_z)/2 \\
\end{gather}
connect site A to its six nearest neighbors
of site B along $\pm\hat{\v{x}}$, $\pm\hat{\v{y}}$ and $\pm\hat{\v{z}}$ directions, respectively.  

In terms of the basis vectors $\v{a}_1$, $\v{a}_2$ and $\v{a}_3$, 
the above hopping terms translate into coupling within the same unit cell
$H_{AB}^{0} = T_{-z}$, and various of hopping terms between neighboring unit cells 
$H_{AB}^{\v{a}_1}= T_x$, $H_{AB}^{\v{a}_1 - \v{a}_3} = T_{-x}$, $H_{AB}^{\v{a}_2} = T_y$, $H_{AB}^{\v{a}_2 - \v{a}_3}=T_{-y}$,
and $H_{AB}^{\v{a}_3} = T_z$, where $H_{AB}^{\v{d}}$ denotes the vector $\v{d}$ denotes the relative position of the involved two unit cells. 

Let $\v{k}$ be the Bloch momenta and $k_j = \v{k}\cdot\v{a}_j$ ($j=1,2,3$), and let us choose $V=M\sigma_z$, then the
Bloch Hamiltonian of this AFMTI can be written as
\begin{equation}
  \mathcal{H}(\v{k}) = \mathcal{H}_0(\v{k}) + V\mu_z,
  \label{eq:bloch_ham_AFMTI}
\end{equation}
where the time-reversal invariant part is
\begin{equation}
	\mathcal{H}_0(\v{k}) = m\rho_z + \mathcal{V}(\v{k})(\cos(k_3/2)\mu_x + \sin(k_3/2)\mu_y),
\end{equation}
with
\begin{align}
	&\mathcal{V}(\v{k}) = t\left[\cos(k_1 - \frac{k_3}{2}) + \cos(k_2 - \frac{k_3}{2}) + \cos(\frac{k_3}{2})\right]\rho_z
  \nonumber
  \\
	&- \lambda \left[\sin(k_1 -\frac{k_3}{2})\sigma_x + \sin(k_2 - \frac{k_3}{2})\sigma_y +
  \sin\frac{k_3}{2}\sigma_{z}\right]\rho_x.
\end{align}
Here we have introduced Pauli matrices $\mu_i, i=x,y,z$ for the sublattice degree of freedom.

Note that $\mathcal{H}(\v{k})$ breaks the time-reversal symmetry because of the term $V\mu_z$. However,
the system is invariant under the composite operation consisting
both a half-period translation along $\v{a}_3$ and the time-reversal operation. 
Formally, we have 
\begin{equation}
S(\v{k})\mathcal{H}(\v{k}) = \mathcal{H}(-\v{k})^*S(\v{k}),
\end{equation}
with $S(\v{k}) = \Theta T_{1/2}(\v{k})$, where 
\begin{equation}
T_{1/2}(\v{k}) = e^{ik_3/2}\left[\cos(k_3/2)\mu_x + \sin(k_3/2)\mu_y\right]
\end{equation}
describes the basis transformation
when the system is translated along $\v{a}_3$ by half a period.

At $k_3 = 0$, an effective time-reversal symmetry realized by $S = -i\mu_x\sigma_y\mathcal{K}$ 
emerges, with $S^{2}=-1$, for the two dimensional Hamiltonian
$\mathcal{H}_{\rm eff}(\v{k}_{12}) =\mathcal{H}(\v{k}_{12},k_3=0)$, such that
\begin{equation}
S\mathcal{H}_{\rm eff}(\v{k}_{12}) = \mathcal{H}_{\rm eff}(-\v{k}_{12})^*S.
\label{eq:effective_TR}
\end{equation}
Note that there is no topological invariants associated with $k_{3}=\pi$, since $S(k_3=\pi)^2=1$. 

\subsection{Band structure of the surface states}
\begin{figure}[t]
	\centering
	\includegraphics[width=0.48\textwidth]{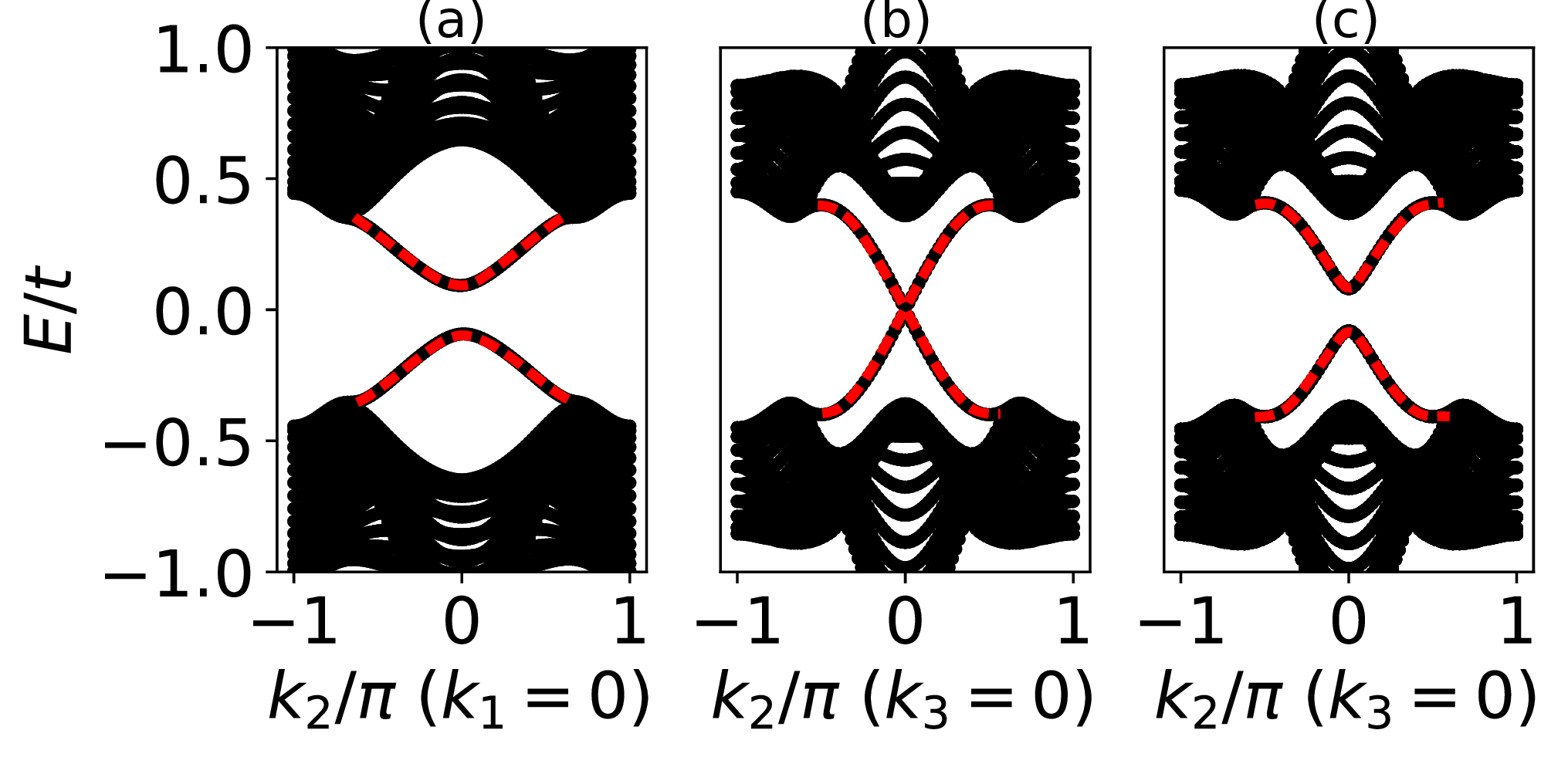}
	\caption{\label{fig:surfaces}Bulk and surface band structures for AFMTI,
		along the (a) $(\bar{1}\bar{1}1)$ and (b,c) $(100)$ surfaces, 
		in which we fixed one momentum ($k_1$ or $k_3$) to zero.
		(c) is the BdG band structure when superconducting pairing potential $\Delta/t=0.1$ was introduced at the surfaces, which
		decays exponentially into the bulk along $\v{a}_1$ with decaying length $\xi=3$.
		The surface states are indicated by red dashed lines.
		The other parameters are $\lambda/t = 0.5$, $m/t = 2$, $M/t=1.2$ with 18 unit cells along the finite direction.  
}
\end{figure}

In this tight-binding model, the (100) surface parallel to $\v{a}_2,\v{a}_3$
is gapless and is of type A (same to the (010) surface), 
whereas the $(\bar{1}\bar{1}1)$ surface parallel to $\v{a}_1,\v{a}_2$ is gapped, and thus is of type F.
The bulk and surface band structures along these terminations are shown in Figs.~\ref{fig:surfaces}(a) and (b),
where we assumed the open boundary condition along $\v{a}_1$, 
and we find that the surface states of type F and type A surfaces are indeed gapped and gapless, respectively.

To numerically confirm the superconducting proximity effect,
we consider a finite number of unit cells along $\v{a}_1$, and choose the periodic boundary condition along $\v{a}_2$
and $\v{a}_3$, such that
the momenta $k_2$ and $k_3$ are still well defined. Using Eq.~(\ref{eq:BdG}),
and taking a spatial dependent pairing potential $\Delta(j) = \Delta \exp(-j/\xi)$ decaying at length scale $\xi$,
the BdG bulk and surface spectra along $(100)$ planes can be calculated. The dispersion of the gapped
surface states at $k_3=0$ are shown in Fig.~\ref{fig:surfaces}(c).

\section{Majorana hinge states \label{sec:MHM}}
Recall that CMMs appear at the domain wall between the two gapped regions,
due to superconductivity and magnetisim, respectively \cite{Fu2008}.
In the proximity coupled AFMTI model introduced above, the common shared hinges between type F and type A
surfaces are exactly such domain walls, as one introduces superconductivity on these type A surfaces.

To demonstrate such chiral Majorana hinge modes, let us assume the
system is finite along $\v{a}_1$ and $\v{a}_3$, and periodic along $\v{a}_2$.
The superconducting proximity effect is modeled by introducing the 
intraorbital $s$-wave pairing potential which decays exponentially from
the $(100)$ surfaces into the bulk, described previously. 

\begin{figure}[t]
	\centering
	\includegraphics[width=0.48\textwidth]{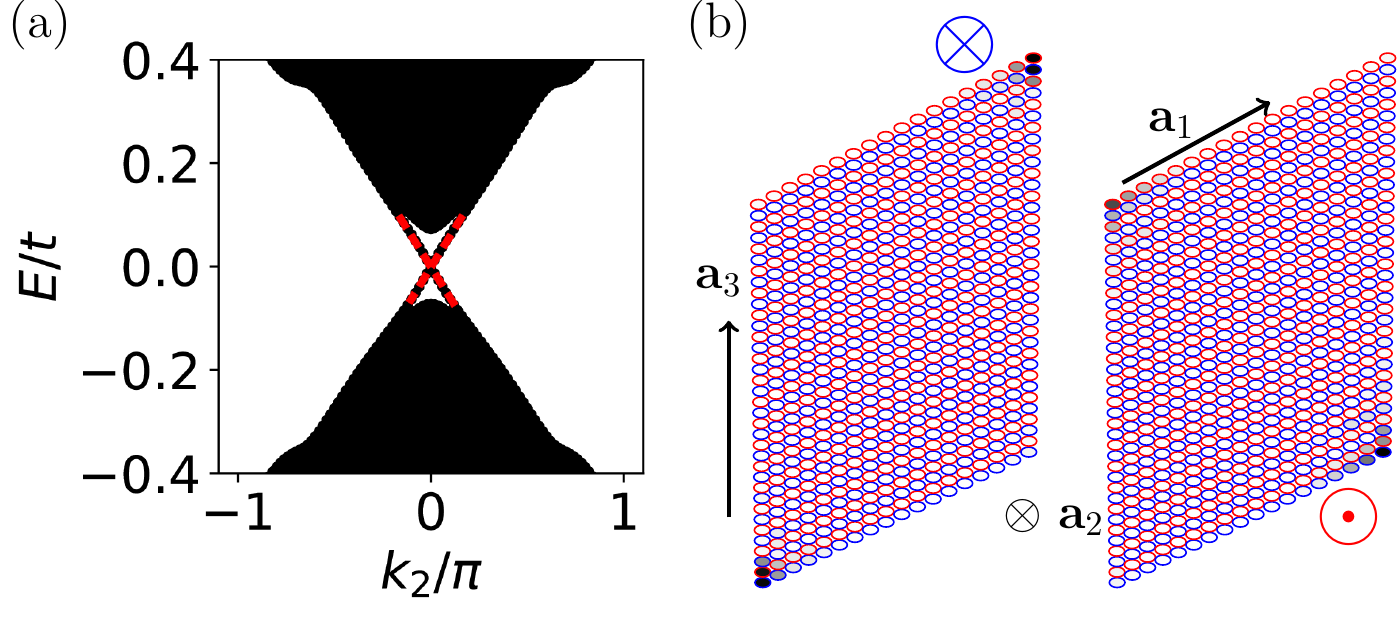}
	\caption{\label{fig:CMM-AF} (a) Bulk and hinge band structure for the proximity coupled AFMTI,
		with periodic boundary condition along $\v{a}_2$ ($k_2$ is well defined).
		The other parameters are the same as in Fig.~\ref{fig:surfaces}(c).
		The hinge states are indicated by red dashed lines. 
		(b) Norm of the hinge state wave functions at $k_2=0$ as a function of positions. 
		The blue and red circles denote the $A$ and $B$ sites. 
		The blackness inside the circles indicate the magnitude of the wave function norm.
		Left: One of the doubly degenerate CMMs propagating inward, along $\v{a}_2$. Right: One of the doubly degenerate CMMs propagating outward, along $-\v{a}_2$.
		The numbers of layers along $\v{a}_1$ and $\v{a}_3$ are 18 and 36 (18 for A and 18 for B),
		respectively.
}
\end{figure}

In Fig.~\ref{fig:CMM-AF}(a), we show the bulk and hinge band structure, 
in which there are doubly gapless chiral modes propagating 
with positive and negative velocities, as indicated
by red dashed lines. These gapless chiral states
are indeed localized around the hinges shared by
$(\bar{1}\bar{1}1)$ and $(100)$ surfaces, 
which are the top/bottom and left/right edges in Fig.~\ref{fig:CMM-AF}(b).

Note that when we have an even number of layers along $\v{a}_3$, the top and 
bottom $(\bar{1}\bar{1}1)$ surfaces will carry opposite magnetization, 
which creates two CMMs with the same chirality located in a diagonal fashion
with respect to each other, as shown in Fig.~\ref{fig:CMM-AF}(b). 
The double degeneracy of the gapless modes is due to
the two fold rotation symmetry with axis along $\v{a}_2$,
which relates the two diagonally aligned hinges along $\v{a}_2$.

\begin{figure}[t]
	\centering
	\includegraphics[width=0.48\textwidth]{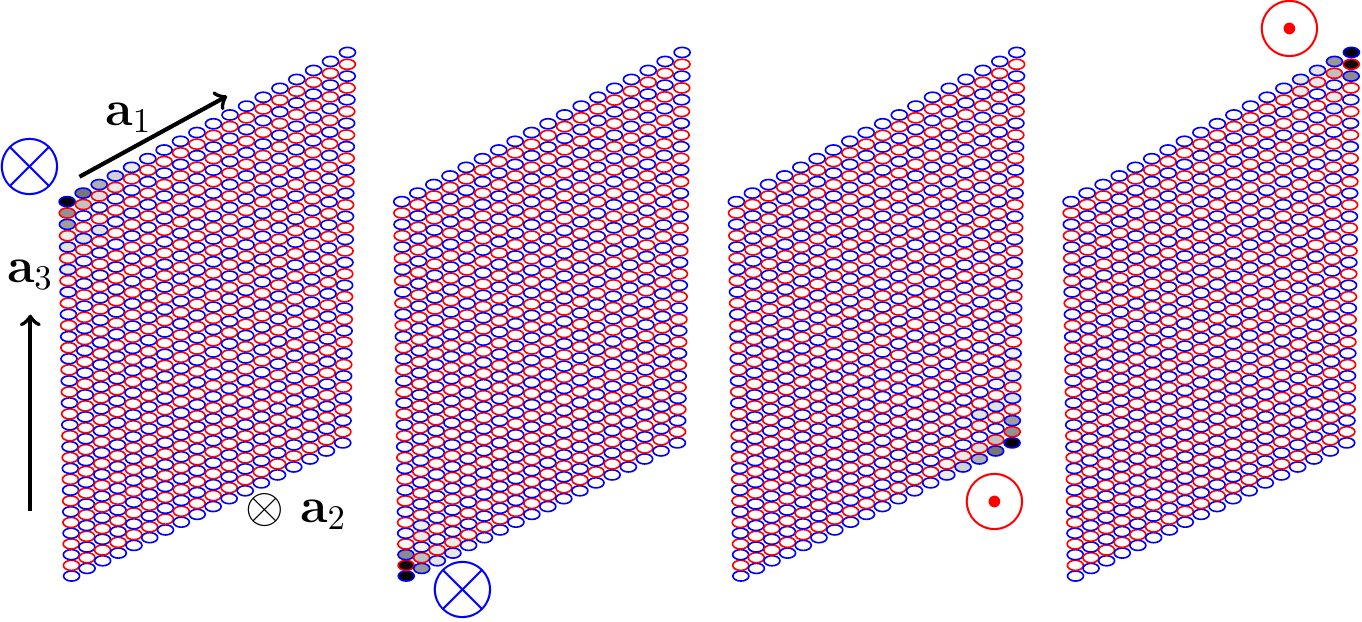}
	\caption{\label{fig:CMM-FM} Norm of the four hinge state wave functions at $k_2=0$ as a function of positions. 
		The blue and red circles denote the $A$ and $B$ sites. 
		The blackness inside the circles indicate the magnitude of the wave function norm.
		The propagating directions of the modes are indicated by the symbol ``$\otimes$''  and ``$\odot$'' for along $\v{a}_2$ and $-\v{a}_2$.
		The numbers of layers along $\v{a}_1$ and $\v{a}_3$ are 18 and 35 (18 for A and 17 for B),
		respectively.
}
\end{figure}

When we change the number of layers along $\v{a}_3$ from even to odd, 
the magnetization of top surface and bottom $(\bar{1}\bar{1}1)$ surfaces 
points to the same direction. We still have four gapless CMMs, due to 
the four edges shared by $(\bar{1}\bar{1}1)$ and $(100)$ surfaces. 
However, the two hinge modes with the same chirality will appear
on the same side of the $(100)$ surface, as shown in Fig.~\ref{fig:CMM-FM}.
Moreover, the two chiral CMMs with the same chirality will have slightly
different velocities, due to lack of symmetry which relates one another.

Thus, by changing the number of layers along the antiferromagnetic order direction
of the AFMTI, one is able to engineer the CMMs with desired propagating directions, 
which can be used, for example, to design a transport experiment detecting
the CMMs, as discussed in the following.

\section{Experimental signature \label{sec:experiments}}
\begin{figure}[t]
	\centering
	\includegraphics[width=0.48\textwidth]{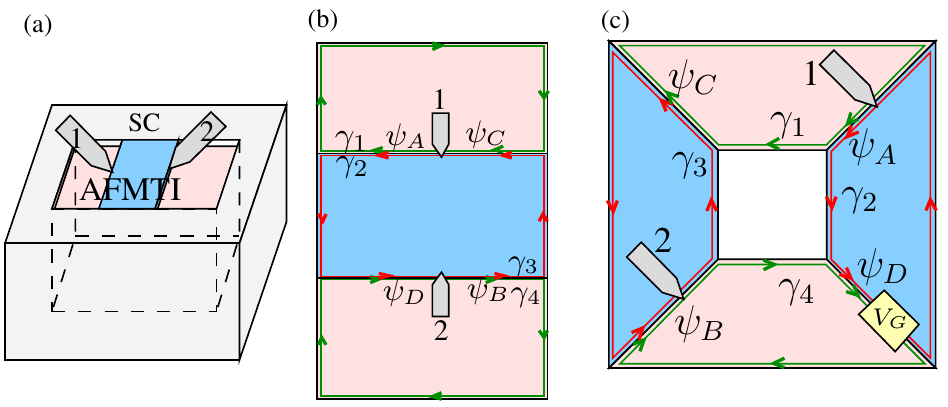}
	\caption{\label{fig:signature} (a) Setup for transport measurement of the CMMs. 
		The top surface of the AFMTI is type F, in which there
		is a region with opposite magnetization direction compared with
		the rest of the surface, as indicated with different colors. 
                All type A surfaces sharing edges with the top surface 
		are in proximity with the superconductor from the side. 
		We connect leads 1,2 at the two domain walls between regions with opposite magnetizations.
		(b) The pattern of the chiral electron modes (double lines) and the CMMs (single lines)
		on the top surface, in which the arrows indicate propagating directions. 
		(c) Setup up for measuring quantum coherence of CMMs, 
		viewed from the top surface, in which the color indicates 
		the magnetization direction. In the yellow region, a gate with
		voltage $V_G$ is added. 	
		The superconductor is proximity coupled 
		to the outer and inner surfaces of the AFMTI 
		from the side. }
\end{figure}

To detect the CMMs at the hinges of the AFMTI, we propose a
transport measurement based on the setup shown in 
Fig.~\ref{fig:signature}(a), in which the AFMTI
is surrounded by the $s$-wave superconductor, 
such that the top surface be type F, and all type A 
surfaces sharing edges with the top surface 
are in proximity with the superconductor. 
Moreover, we require the top surface to have
a region in which the number of layers
along the $T_{1/2}$ direction differs by one
from that of the rest of the surface. 
This type of device may be created by 
taking a Nb superconductor which is hollow inside.
Thus, one can use it as a mask to 
grow the layered $\mathrm{MnBi_2Te_4}$ inside of the
superconductor. 

Because of the antiferromagnetic ordering, 
this creates two domain walls between regions 
with different magnetizations on the top surface.
Thus, we expect to have a single chiral electron modes
on each of the domain walls, propagating in opposite directions \cite{Mong2010, Zhang2013, Li2018},
see App.~\ref{app:domain-wall} for details.
Furthermore, there are CMMs appearing on these sharing edges 
between the two types of surfaces. The propagating directions of these CMMs
are determined by the magnetization direction, and
the relative alignment between type F and type A surfaces. 

In Fig.~\ref{fig:signature}(b), we illustrate these chiral electron and
Majorana modes in green and red lines on the edges of the top surface, 
with arrows indicating the their propagating directions. 
Note that the chiral electron mode are shown as (green and red) double lines 
given the fact that the chiral electron mode can be decomposed as two chiral CMMs.
We further connect leads 1 and 2 to these two chiral electron modes, as illustrated
in the figure, and measure the conductance $\sigma_{12}$ between them.
We show in the following that $\sigma_{12} = \frac{e^2}{2h}$, same as the 
signature of CMM proposed in the QAHI-TSC-QAHI system \cite{Chung2011,He2017}.

Let us denote the chiral electron mode flowing out from (into) leads 1 and 2
as $\psi_{A}$ and $\psi_B$ ($\psi_{C}$ and $\psi_{D}$). These modes can be decomposed
into CMMs as $\psi_A = (\gamma_1 + i\gamma_2)/2$, $\psi_{B} = (\gamma_4 + i\gamma_3)/2$, 
$\psi_{C} =(\gamma_1 - i\gamma_3)/2$, $\psi_D = (\gamma_4+i\gamma_2)/2$ \cite{Chung2011}. 
Consider a scattering event by regarding $(\psi_A, \psi_{A}^\dagger, \psi_{B},\psi_{B}^\dagger)$ as 
incident modes, and $(\psi_C, \psi_C^\dagger, \psi_D, \psi_{D}^\dagger)$ as outgoing modes,
then the scattering matrix $\mathbf{S}$, which relates the incident modes and the outgoing modes, 
can be obtained, see App.~\ref{app:exp-signature}. 
In particular, we find the probabilities for an incident electron from lead $1$ 
in channel $\psi_A$ transmits into $\psi_C$ as an electron and into $\psi_C^\dagger$ as
a hole are both $1/4$, which leads to the two-terminal conductance $\sigma_{12}=\frac{e^2}{2h}$
according to the generalized Landauer formula \cite{Takane1992}.

The quantum coherence of the CMMs can be demonstrated
using an Majorana interferometer depicted in Fig.~\ref{fig:signature}(c),
in which we add a gate at voltage $V_G$ in a region of the chiral 
electron mode $\psi_D = \gamma_4+i\gamma_2$, 
creating a term $H_G =V_G\psi_D^\dagger\psi_D$
within a length $l_G$ through which $\psi_D$ travels. 
This leads to a phase-dependent two-terminal conductance $\sigma_{12} = (1+\cos\varphi_G)e^2/2h$ \cite{Lian2017}.

\section{Conclusions \label{sec:conclusion}}
In this work, we proposed to realize CMMs
in a 3D AFMTI (such as $\mathrm{MnBi_2Te_4}$) in proximity with an $s$-wave superconductor (such as Nb).
This is based on an important result of our work, 
that the type A surface of an AFMTI can indeed be gapped out 
by proximity-induced $s$-wave superconductivity.
More interestingly, the nature of this gapped surface state
is the same as the one from gapping out a TRITI surface with superconductivity, 
despite the absence the physical time-reversal symmetry.

Our proposal has certain advantages over the existing platform for CMMs.
First, since the material realization of the AFMTI, the 
$\mathrm{MnBi_2Te_4}$-related ternary chalgogenides, are 3D bulk crystals
with intrinsic magnetic order, the temperature for observing the CMMs
will be presumably higher than the one in the existing 2D CMMs platforms, 
such as Cr-doped $\mathrm{(Bi,Sb)_2Te_3}$, which is extremely inhomogeneous.
Second, as there is no need to introduce external magnetic field,
which is required in the 2D platforms, the complication when the magnetism and superconductivity are spatially overlapping
can be avoided.

Another nice feature of the proposed system is that 
a new degree of freedom, namely the number of layers, emerges and can be manipulated.
We have demonstrated that one is able to create a network of chiral propagating electron modes and CMMs
in a controlled fashion~\cite{Mong2010,Li2018},
by engineering step edges on type F surfaces
of the AFMTI. This can be applied to the detection the CMMs,
in terms of a measurement of the two-terminal conductance. 
Moreover, 
a Majorana interferometer in Fig.~\ref{fig:signature} can also be 
created, which will demonstrate the braiding properties of the CMMs \cite{Lian2017}.
Thus, the AFMTI/superconductor platform is an excellent candidate for 
topological quantum computing with CMMs. 

Last but not least, our proposal realizes an extrinsic 3D second-order topological superconductor, 
making our work also valuable to the active field of searching for topological corner and hinge modes. 
This platform may be used to study the difference between the CMMs in 3D and the ones 2D,
such as the spatial localization behavior,
as well as the fate of the CMMs under additional perturbations \cite{Queiroz2018}. 

\acknowledgments
Y.P. acknowledges support from the IQIM,
an NSF physics frontier center funded in part by the Moore Foundation, 
and support from the Walter Burke Institute for Theoretical Physics at Caltech. 
Y.X. acknowledges support from the Basic Science Center Project of NSFC (Grant No. 51788104), the Ministry of Science and Technology of China (Grants No. 2018YFA0307100 and No. 2018YFA0305603), the National Thousand-Young-Talents Program and Tsinghua University Initiative Scientific Research Program.

\appendix
\section{Chiral electron modes on the domain wall of the AFMTI} \label{app:domain-wall}
Due to the antiferromagnetic order in the AFMTI, the magnetization direction of type F surface alternates
as we change the number of layers along the antiferromagnetic direction. 
The step edge between regions with layer numbers differ by one can be regarded as a domain wall,
on which the massive Dirac field changes its sign. Thus, a chiral electron mode 
is expected on this step edge \cite{Mong2010}. In the following, we show this
chiral electron mode in the tight-binding AFMTI model, as well as the coexisting chiral
Majorana modes.

\begin{figure*}
	\centering
	\includegraphics[width=0.9\textwidth]{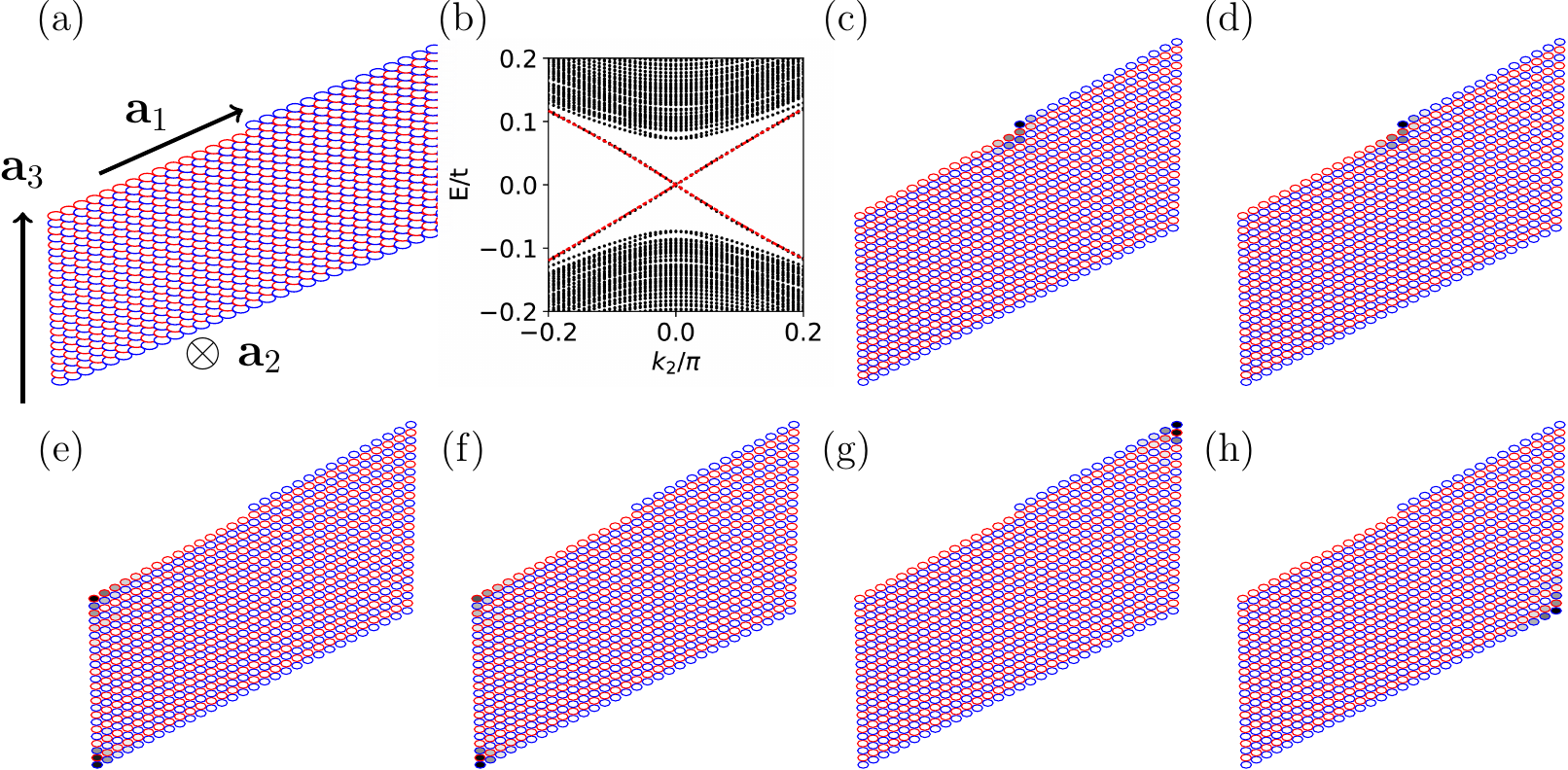}
	\caption{\label{fig:step} AFMTI with a step edge along $\v{a}_2$ on one of the type F surface parallel to $\v{a}_1$ and $\v{a}_2$.
	The system is finite the directions of $a_1$ and $a_3$, and periodic in $\v{a}_2$. The blue and red circles denote the
	A and B sites. (b) Bulk and hinge (including the step edge) band structure for the AFMTI, 
	in which the left and right surfaces parallel to $\v{a}_2$ and $\v{a}_3$ are gapped out by proximity induced superconducivity,
	with a pairing potential exponentially decaying into the bulk. The gapless hinge states (electron or Majorana modes)
	are indicated in red. (c--h) Norm of the hinge state wave functions at $k_2=0$ as a function of positions. 
	The blackness inside the circles indicate the magnitude of the wave function norm.
	(c,d) Chiral electron modes (doubled in BdG Hamiltonian)  on the step edge.
(e-h) Four CMMs on the four outer hinges. The parameters are $\Delta/t=0.1$,
$\lambda/t=0.5$, $m/t = 2$, $M/t = 1.2$, $\xi=3$. The system contains 
$30$ layers along $\v{a}_1$, and 24 or 25 layers along $\v{a}_3$. }
\end{figure*}

Let us take the previous introduced AFMTI model, and assume the system is finite along
$a_1$ and $a_3$, and periodic along $\v{a}_2$. We further assume there is a step edge 
along $\v{a}_2$, on one of the type F surface parallel to $\v{a}_1$ and $\v{a}_2$, as shown in 
Fig.~\ref{fig:step}(a). The proximity induced superconductivity is introduced by hand
by adding a pairing potential $\Delta$, which decays exponentially into the bulk at a length scale $\xi$,
on the left and right surfaces parallel to $\v{a}_2$ and $\v{a}_3$.

Since the system is periodic along $\v{a}_2$, one can 
go the momentum space and compute the BdG band structure as a function of the corresponding momenta
$k_2$, as shown in Fig.~\ref{fig:step}(b). We actually obtain six chiral modes inside the 
bulk gap. Among these gapless modes, two of them correspond to a chiral electron mode localized
at the step edge, whose wave functions at $k_2=0$ are shown in Figs.~\ref{fig:step}(c,d). The rest four are chiral Majorana
modes localized at the outer four hinges of the AFMTI, with wave functions at $k_2=0$ shown in Figs.~\ref{fig:step}(e--h).

\section{Signatures of CMMs in terms of two-terminal conductance $\sigma_{12}$} \label{app:exp-signature}
In this section, we provide more details on the calculation of the two-terminal  conductance $\sigma_{12}$.

Let us first consider the setup in Fig.5(b) of the main text.
One can imagine the system as a normal-superconductor-normal junction 
with chiral electron modes flowing out from (into) leads 1 and 2
are $\psi_{A}$ and $\psi_B$ ($\psi_{C}$ and $\psi_{D}$), which can be decomposed
into CMMs as $\psi_A = (\gamma_1 + i\gamma_2)/2$, $\psi_{B} = (\gamma_4 + i\gamma_3)/2$,
$\psi_{C} =(\gamma_1 - i\gamma_3)/2$, $\psi_D = (\gamma_4+i\gamma_2)/2$ \cite{Chung2011}.
Because of this decomposition, we have 
\begin{equation}
	\left(\begin{array}{c}
\psi_{C}\\
\psi_{C}^{\dagger}\\
\psi_{D}\\
\psi_{D}^{\dagger}
\end{array}\right)= \mathbf{S} \left(\begin{array}{c}
\psi_{A}\\
\psi_{A}^{\dagger}\\
\psi_{B}\\
\psi_{B}^{\dagger}
\end{array}\right),\quad 
	\mathbf{S} =\frac{1}{2}\left(\begin{array}{cccc}
1 & 1 & -1 & 1\\
1 & 1 & 1 & -1\\
1 & -1 & 1 & 1\\
-1 & 1 & 1 & 1
\end{array}\right),
\end{equation}
where $\mathbf{S}$ is the scattering matrix. 

The two-terminal conductance is given by generlized Landauer formula \cite{Takane1992} 
\begin{equation}
	\sigma_{12} = \frac{g_{11}g_{22}-g_{12}g_{21}}{g_{11}+g_{22}+g_{12}+g_{21}} 
\end{equation}
where
\begin{equation}
	g_{ij} = \frac{e^2}{h}(\delta_{ij} - \abs{S_{ij}^{ee}}^2 + \abs{S_{ij}^{eh}}^2 ),
\end{equation}
with $i,j=1,2$ corresponding to the lead label, and $S_{ij}^{\alpha\beta}$ ($\alpha,\beta=e,h$)
is the matrix element of $\mathbf{S}$, in which the basis is ordered as $(1e,1h,2e,2h)$.
Using the scattering matrix $\mathbf{S}$, we have $g_{11}=g_{22}=e^2/h$ and $g_{12}=g_{21}=0$.
Thus, $\sigma_{12} = e^2/2h$.

This conductance can also be obtained in the following way \cite{Lian2017}. 
Let us use $\ket{n_X n_Y}$ to denote an eigenstate of the occupation operators of modes $\psi_{X},\psi_{Y}$, with eigenvalues $n_X,n_Y$ ($X,Y=A,B,C,D$).
If we consider one electron comming from from lead 1 or 2, then the system is prepared in state $\ket{1_A 0_B}$ or $\ket{0_A1_B}$, 
which translate into a linear combination of the basis state in the outgoing channel via
\begin{equation}
	\left(\begin{array}{c}
\ket{1_{C}0_{D}}\\
\ket{0_{C}1_{D}}
\end{array}\right)=\mathcal{M}\left(\begin{array}{c}
\ket{1_{A}0_{B}}\\
\ket{0_{A}1_{B}}
\end{array}\right),
\quad
\mathcal{M} = 
\frac{1}{\sqrt{2}}\left(\begin{array}{cc}
1 & -1\\
1 & 1
\end{array}\right)
\end{equation}
Hence, we see the probability of finding the electron incident from lead 1 or 2
is given by $\abs{\braket{0_C 1_D|1_A0_B}}^2=1/2$, giving rise to
the conductance of $e^2/2h$.

To compute the two-terminal conductance $\sigma_{12}$ in Fig.5(c) of 
the main text, we make use of the above approach by considering an incident electron coming from lead 1 or 2,
namely we prepare the system in $\ket{1_A 0_B}$ or $\ket{0_A1_B}$.
We can imagine the electron propagation in this system in terms of two steps.
First, the incident electron propagates into modes $\psi_C$ and $\psi_D$ (left top and right bottom of Fig.5(c))
after experiencing the the gate voltage $V_G$, which transforms $\ket{0_C1_D} \to e^{-i\varphi_G}\ket{0_C 1_D}$.
In the second step, the electron propagates back into mode $\psi_A'$ and $\psi_B'$,
where $\psi_A'$ and $\psi_{B}'$ denote the outgoing mode into leads 1 and 2.
The whole process can be described by the following transformation
\begin{equation}
	\left(\begin{array}{c}
\ket{1_{A'}0_{B'}}\\
\ket{0_{A'}1_{B'}}
\end{array}\right)=\mathcal{M}\mathcal{V}\mathcal{M}\left(\begin{array}{c}
\ket{1_{A}0_{B}}\\
\ket{0_{A}1_{B}}
\end{array}\right),\quad\mathcal{V}=\left(\begin{array}{cc}
1 & 0\\
0 & e^{-i\varphi_{G}}
\end{array}\right).
\end{equation}
The conductance is thus given by 
\begin{equation}
	\sigma_{12}=\frac{e^2}{h}\abs{\braket{0_{A'}1_{B'}|1_A0_B}}^2  = \frac{1+\cos\varphi_G}{2}\frac{e^2}{h}.
\end{equation}

\end{document}